\documentclass[fleqn,twoside]{article}

\usepackage{espcrc2}
\usepackage{amsmath} 
\usepackage{amssymb} 
\usepackage{graphicx} 
\usepackage{tabularx}

\newcommand{\C}{\mathbb{C}}
\newcommand{\R}{\mathbb{R}}
\newcommand{\N}{\mathbb{N}}
\renewcommand{\r}[1]{\mathrm{#1}}

\newcommand{\fig}[2]{\includegraphics[width=#2\textwidth]{#1}}

\title{Multiple return times in the quantum baker map} 
\author{M.~Fannes\address[L]{Instituut voor Theoretische Fysica, Katholieke
Universiteit Leuven,\\ Celestijnenlaan 200D, B-3001 Heverlee,
Belgium}\thanks{Email: mark.fannes@fys.kuleuven.ac.be},
P.~Spincemaille\addressmark\thanks{Email: pascal.spincemaille@fys.kuleuven.ac.be}\thanks{Acknowledges financial support from FWO project G.0239.96}}

\begin{document}

\begin{abstract}
For chaotic classical systems, the distribution of return times to a small
region of phase space is universal.  We propose a simple tool to investigate
multiple returns in quantum systems.  Numerical evidence for the baker map and
kicked top points, also in the quantum case, at a universal distribution.\\
{\it PACS} : 05.45.Mt,05.45.Pq \\ 
{\it keywords} : return times, baker
map, kicked top
\end{abstract}

\maketitle

\section{Introduction}

In classical dynamical systems, one may study how long it takes a system to
return to the phase space region it occupied at time zero.  In general, this
time will increase when the phase space area under consideration is decreased.
M.~Kac has shown that, for an ergodic dynamical system, the average return
time, assuming that the system starts in a set $A$, is given by $\mu(A)^{-1}$
where the average is taken over the ergodic measure $\mu$. 

Related to this are results about asymptotically rare events. Consider a
sequence $A_k$ of subsets of phase space such that
$\lim_{k\rightarrow\infty}\mu(A_k)=0$. If the time for a point $x$ in phase
space to visit $A_k$ is denoted by $\tau_k(x)$, then the question arises how
the product $\mu(A_k) \tau_k(x)$ behaves. This turns out, for a large class of
models, to have an exponential distribution~\cite{HiSaVa}. 

One may wonder whether a similar situation occurs in quantum mechanics.  In
finite dimensional systems, looking with increasing precision (corresponding to
the $k \rightarrow \infty$ limit of above) means increasing the dimension of
the system, which amounts to taking its classical limit. Here, the inverse
dimension plays the r\^ole of Planck's constant. Return times were studied for
various quantum dynamical systems. A natural way of measuring returns is
provided by the notion of transition probability: for any two vectors $\varphi$ and $\psi$ in
a Hilbert space it is defined as $|\langle\varphi|\psi\rangle|^2$. 

\section{Multiple return times}

If a system starts from an initial state $\varphi_0$, one can consider 
\begin{equation}\label{overlap}
 n\in\N\mapsto\langle\varphi_0|\varphi_n\rangle \in \C,
\end{equation}
where $\varphi_n$ is the state of the system after $n$ time steps. For
finite dimensional systems this function is almost periodic.
Interesting behaviour is displayed only when looking at the correct time
scale. It is known that, for classically chaotic systems, the quantum system
follows its classical counterpart in the sense of coherent states up to a
certain breaking time of the order of $\log N$, where $N$ is the dimension of
its Hilbert space of states. We propose to  analyse the
function~(\ref{overlap}) through the spectrum of the Gram  matrix associated
with a sequence of normalised vectors $\varphi_0, \cdots, \varphi_{K-1}$:
\begin{equation*}
 G_K = 
 \begin{pmatrix}
 \langle\varphi_0|\varphi_0\rangle & \cdots &
 \langle\varphi_0|\varphi_{K-1}\rangle\\
 \vdots & \ddots & \vdots\\
 \langle\varphi_{K-1}|\varphi_0\rangle &\cdots &
 \langle\varphi_{K-1}|\varphi_{K-1}\rangle \\
 \end{pmatrix}.
\end{equation*}
$G_K$ is a positive semi-definite matrix and its eigenvalues are
independent of the order of the vectors and of any extra phase added to
any of them. 

Information about the amount and the frequency of vectors that return
to a small neighbourhood in Hilbert space lies encoded in the spectrum
of this matrix. This fact is best appreciated by examining a simple
example. Consider the sequence
$\{\psi_1,\psi_2,\psi_2,\psi_1,\psi_3,\psi_4,\psi_1\}$, where the
$\psi_i$'s are normalised and mutually orthogonal. This amounts to specifying
a sequence of letters belonging to a given alphabet (the classical case). The
Gram matrix consists of only zeroes and ones (because of the orthonormality)
and its spectrum is $\{0,0,0,1,1,2,3\}$. An eigenvalue different from zero
indicates the number of times a certain vector is repeated and its multiplicity
tells how many vectors are repeated that often. In our example, only two vectors
are repeated, the vector $\psi_1$ twice and $\psi_2$ once, summing up to three,
which is exactly the multiplicity of zero. This example provides the link with
the return times mentioned in the introduction: one considers a partition of
phase space and keeps track of the different sets a particle visits during its
evolution. The associated Gram matrix has a one for its $ij$th entry if the
visited sets at times $i$ and $j$ are the same and a zero otherwise. Hence
calculating the spectrum of the Gram matrix immediately gives all the return
times.  

In general, for a true quantum system, the vectors in a sequence will not be
orthogonal and the overlaps will be complex numbers with absolute value smaller
than or equal to one. However, a similar interpretation of the spectrum can be
obtained remembering that overlaps with absolute value close to 1 indicate
vectors close to each other, while overlaps close to zero indicate almost
orthogonal vectors.  Generically, the non-zero eigenvalues of the Gram matrix
are non-degenerate and the notion of multiplicity of above should be
replaced by the spectral density, i.e. a degenerate eigenvalue is now replaced
by different eigenvalues close to each other. The multiplicity of zero is equal
to the number of linear dependencies in the given sequence.  Two rather extreme
cases will clarify this.  A large concentration of eigenvalues around 1
indicates that a large amount of vectors have small mutual overlaps.  This can
be compared with the classical situation of above, with a spectrum consisting
mainly of ones and only a few zeroes, indicating very few repetitions. A
presence of many eigenvalues close to zero suggests that many vectors lie close
to one another, leading to many large overlaps.  This compares to a classical
situation where some letters are repeated many times, giving rise to many
zeroes in the spectrum and only a few large integers.

A point that has not been stressed yet is the number of time steps to
consider.  The spectrum of the Gram matrix will surely contain an eigenvalue
zero when the number $K$ of vectors exceeds the dimension $N$ of the Hilbert
space. Including too many time steps causes a large degeneracy of zero. We
shall demonstrate, using numerical simulations, that also $K$ has to be taken
large such that the ratio $K/N$ is a fixed number $\tau$: time measured on the
Heisenberg scale.  The same scaling is relevant for a random model that can be
treated analytically. 

A natural choice for the initial state is a coherent state centred around a
point in classical phase space. There are several reasons for this. Taking the
large $N$ limit immediately implies the comparison of quantum systems with
different dimensions. This also entails that initial states for different
dimensions have to be comparable. Coherent states exist for every $N$, occupy a
volume  $1/N$ of phase space and have a clear limit, namely a single point in
phase space. Also, a coherent state is typically not a eigenvector of the
evolution which would lead to a completely trivial spectrum of the Gram matrix.

The precise object we shall investigate is the empirical eigenvalue
distribution of the Gram matrix. If a matrix has eigenvalues
$\lambda_1,\cdots,\lambda_K$, then its empirical eigenvalue distribution is
\begin{equation*}
 \rho(\rm d\lambda) \equiv \sum_{i=1}^K \delta(\lambda-\lambda_i)\, \rm
 d \lambda.
\end{equation*}
In this way, the number of eigenvalues in a set $\Lambda\subset\R$ is
given by $\rho(\Lambda)$.

We shall illustrate this approach by example. We consider two simple quantum
dynamical systems frequently encountered in the quantum chaos literature. They
have clearly defined classical limits. The first example is the quantum baker
map and the second the kicked top.

\section{Models}

The quantum baker map is a dynamics on $\C^N$ with $N$ even~\cite{BaVo}.
Defining the discrete Fourier transform on $\C^M$
\begin{equation*}
 F_M |m\rangle = \frac{1}{M} \sum_{n=0}^{M-1} \mathrm e^{\frac{2\pi i n m}{M}}
 |n\rangle,
\end{equation*}
the unitary evolution operator of the baker map is
\begin{equation*}
 U = F_N 
 \begin{pmatrix}
 F^{-1}_{N/2} & 0 \\
 0 & F^{-1}_{N/2} \\
 \end{pmatrix}.
\end{equation*}
Its classical limit can be shown to be
\begin{equation*}
 \begin{pmatrix}
 q \\
 p \\
 \end{pmatrix}
 \mapsto
 \begin{pmatrix}
 2q \bmod 1 \\
 p/2 + [2q]/2\\
 \end{pmatrix} 
\end{equation*}
where $\begin{pmatrix}q \\p \\ \end{pmatrix}$ is a point in the phase
space $[0,1]\times[0,1]$. The position and momentum $q$ and $p$ are the
classical limits of 
\begin{equation*}
 Q_N |m\rangle = \frac{m}{N} |m\rangle
 \quad\mbox{and}\quad
 P_N = F_N Q_N F_N^{-1}.
\end{equation*} 
The baker map is an example of a chaotic system: it has a positive
Lyapunov exponent equal to $\log 2$.

\begin{figure*}[htb]
\begin{tabularx}{\textwidth}{lccc}
     & $\tau=.5$  & $\tau=1$   & $\tau=1.5$ \\
$N = 500$ & \fig{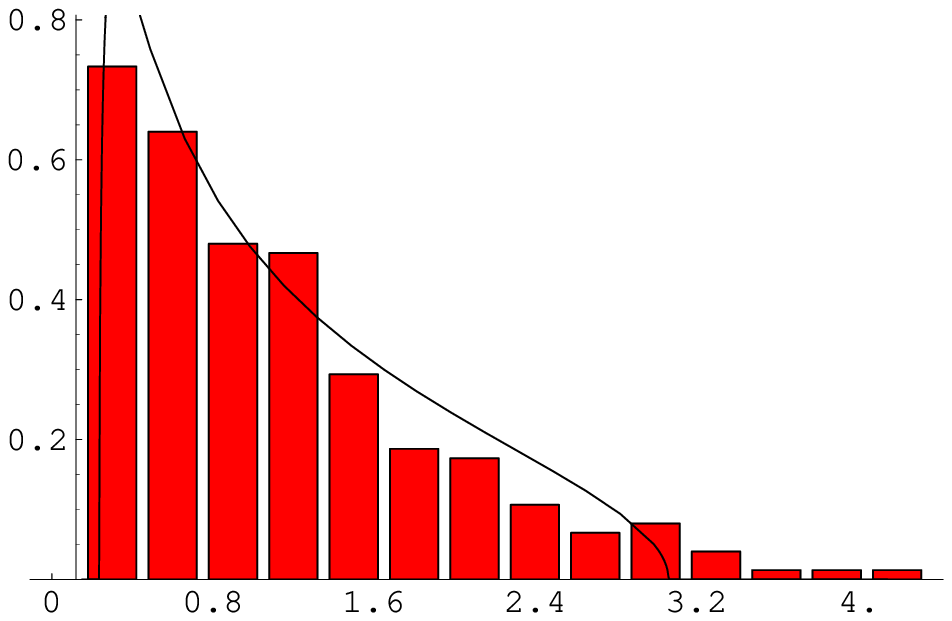}{.25} & \fig{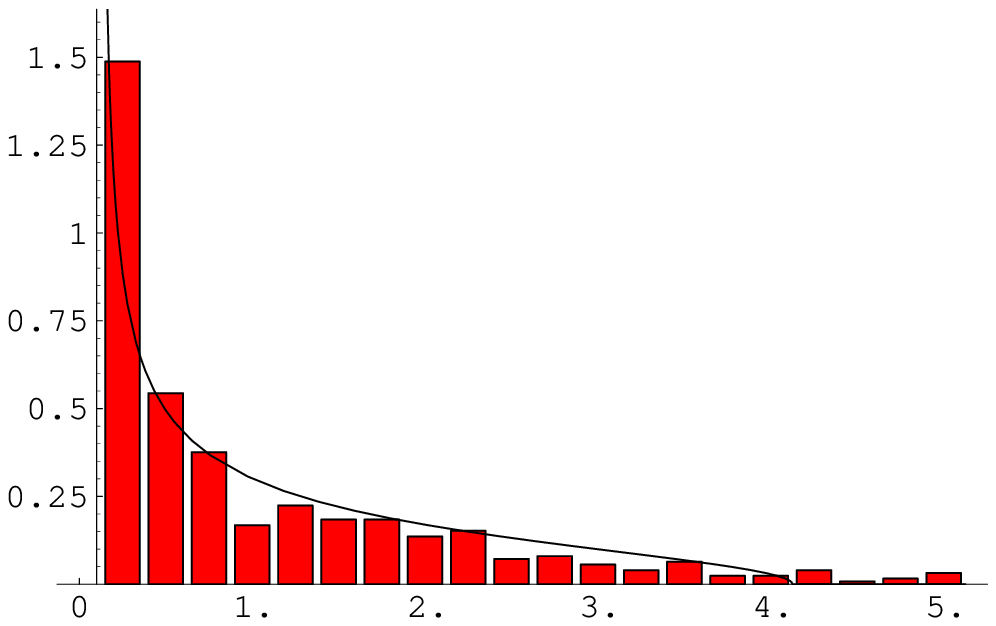}{.25} & \fig{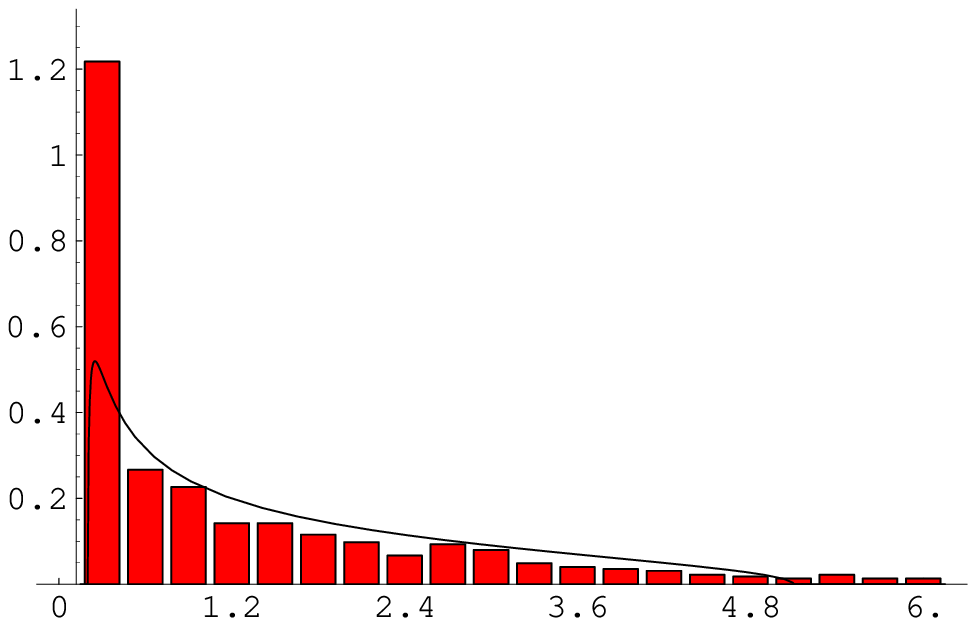}{.25} \\
$N =1000$ & \fig{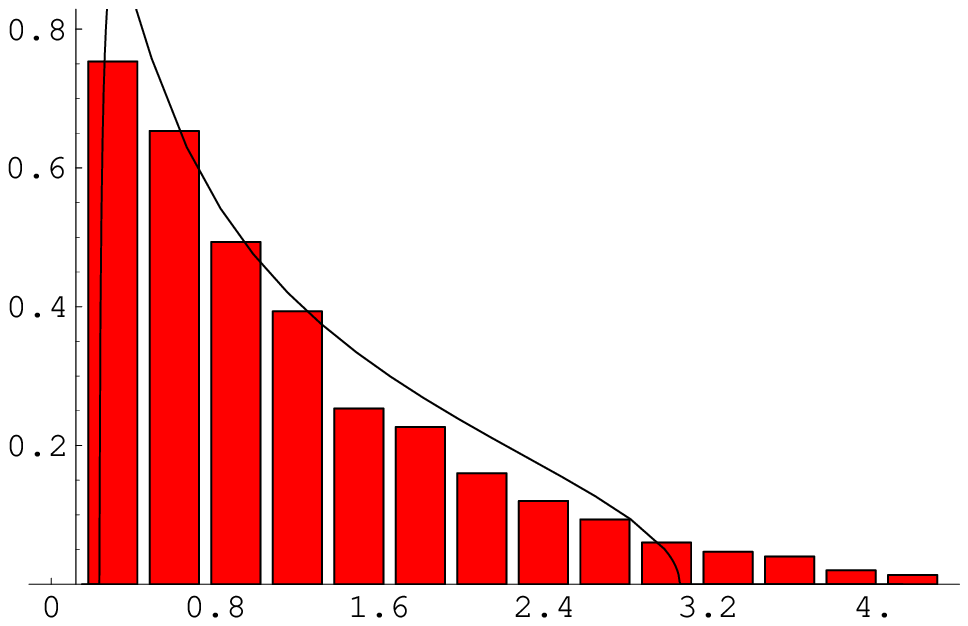}{.25} & \fig{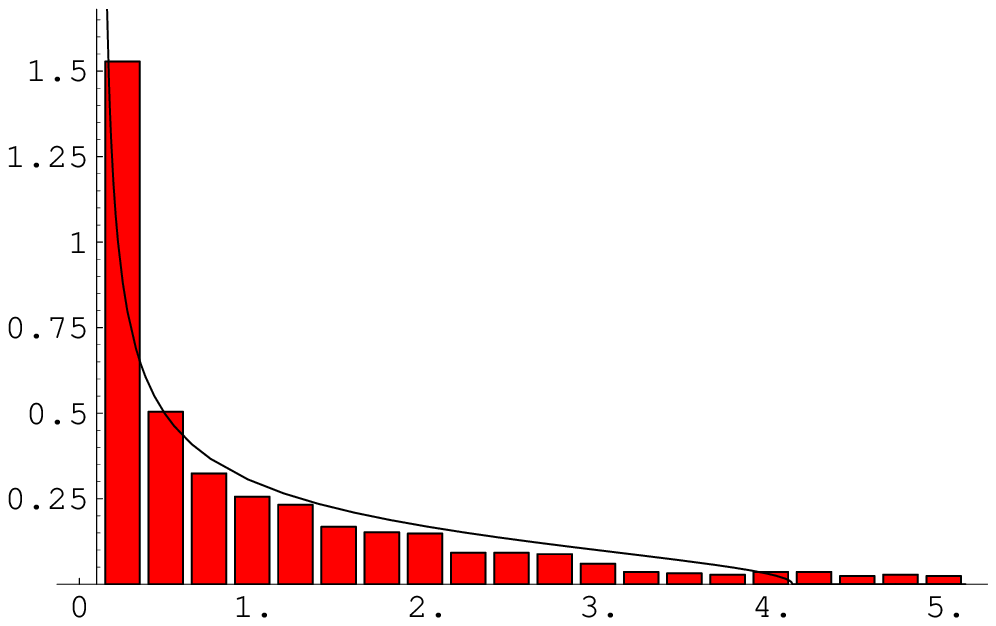}{.25} & \fig{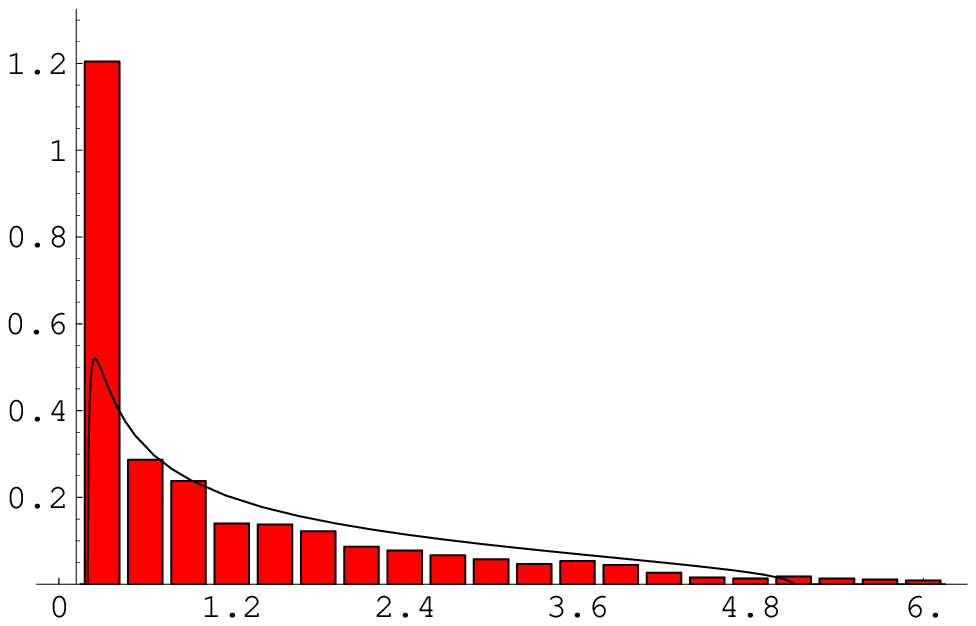}{.25} \\
$N =1500$ & \fig{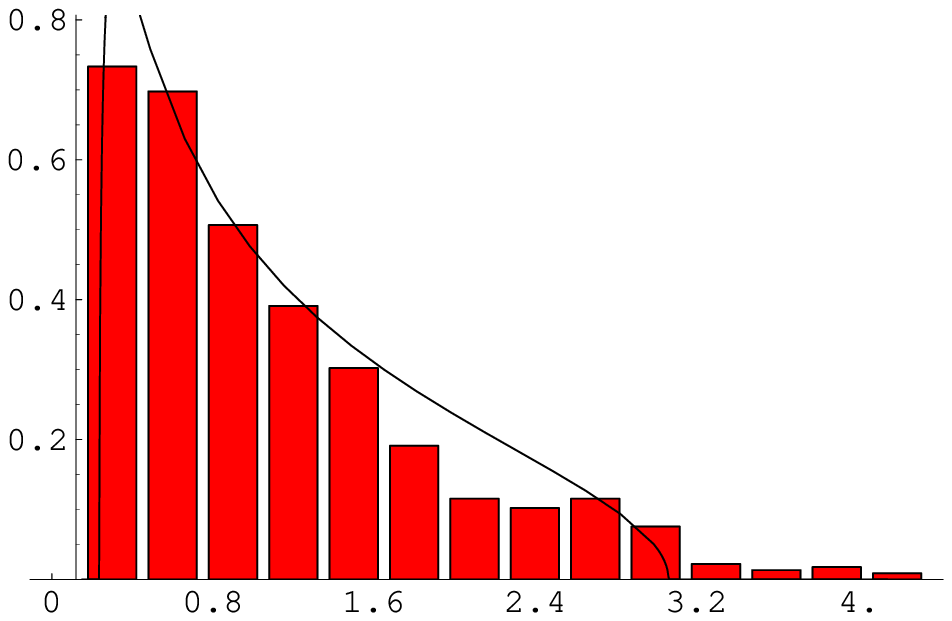}{.25} & \fig{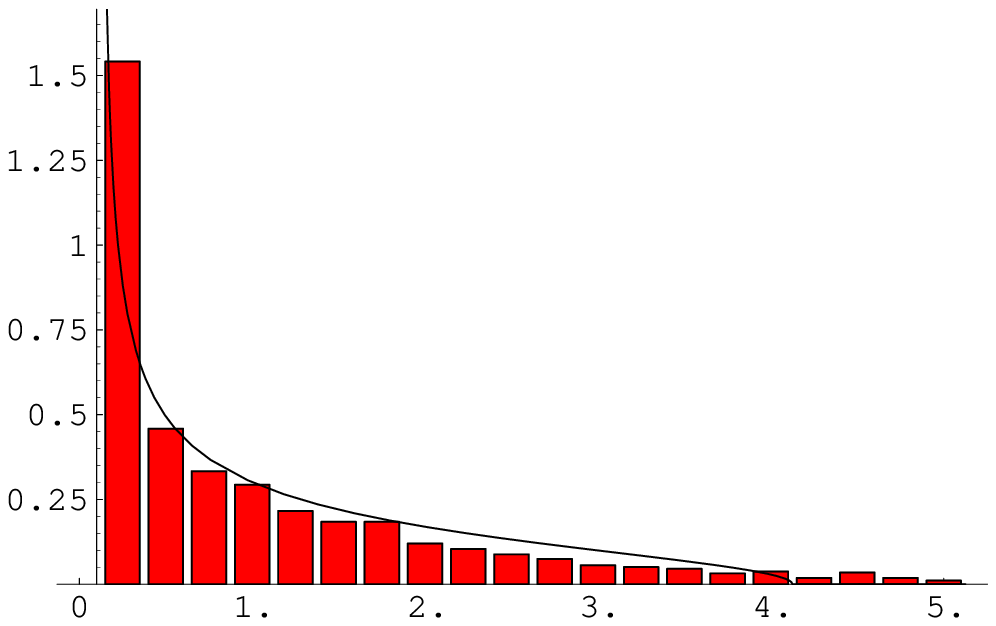}{.25} & \fig{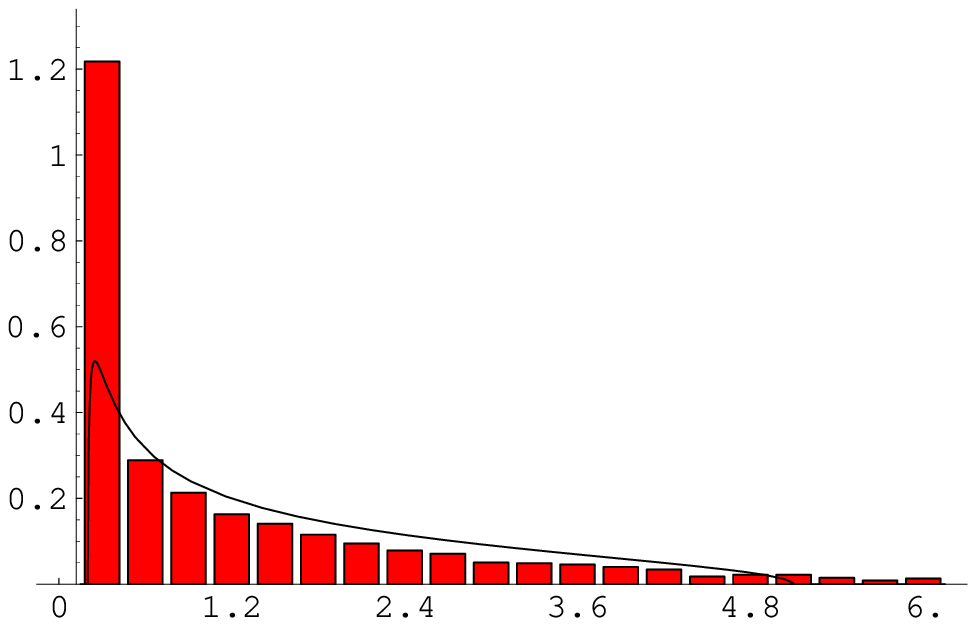}{.25} \\
\end{tabularx}
\caption{Eigenvalue density of Gram matrices for the quantum baker map for
various values of $N$ and $\tau = K/N$} 
\label{bakerfig}
\end{figure*}

First, we numerically construct a coherent state in an $N$-dimensional
Hilbert space centred around a chosen classical point in phase space.
One way of doing this is by calculating the ground state of an
appropriate Harper operator and subsequently shifting it over to the
desired classical point. Then, we apply the operator $U$ a number of
times to calculate the overlaps needed in the Gram matrix. Note that
when the first row of the Gram matrix is calculated, all other matrix
elements are known because of the unitarity of the evolution. Unitarity
has another interesting consequence: the precision of the calculation
does not exponentially decrease with the number of time steps. One way
to see this, is applying the unitary evolution to an initial state many
times, followed by applying an equal number of times the adjoint of the
evolution. The resulting state is remarkably close to the original
state, despite the presence of numerical round-off errors in the
components of the vectors and the unitary matrix. This is in sharp
contrast with numerical simulations of chaotic classical systems where
positive Lyapunov exponents cause the numerical errors to quickly
exceed the desired precision. The eigenvalues of the Gram matrix are
calculated and the empirical eigenvalue distribution is represented by
the bar diagrams of Figure~\ref{bakerfig}. The solid black line will be
commented on later in this letter.

The first thing that Figure~\ref{bakerfig} reveals is the time scale
$N$ on which a non-trivial eigenvalue density is present. For a fixed
time on the Heisenberg time scale, i.e.\ a fixed ratio $\tau = K/N$ , the
eigenvalue densities are very similar.  For comparison, choosing $K$ of the
order of $\log N$ produces a Gram matrix close to the identity matrix and
therefore a spectrum which is greatly concentrated around one. For times much
larger than $N$ the linear dependencies dominate the picture and the empirical
eigenvalue distribution tends to a Dirac distribution at zero.  Secondly, we
note that we did not specify which coherent state we began with.  Our
simulations show that the initial state does not affect the form of the
eigenvalue distribution.  Diagrams, very similar to those of
Figure~\ref{bakerfig}, were obtained for different initial states.

\begin{figure*}[htb]
\begin{tabularx}{\textwidth}{lcc}
          & $\tau=.5$  & $\tau=1$   \\
$k=1.5$ and $p=1$  & \fig{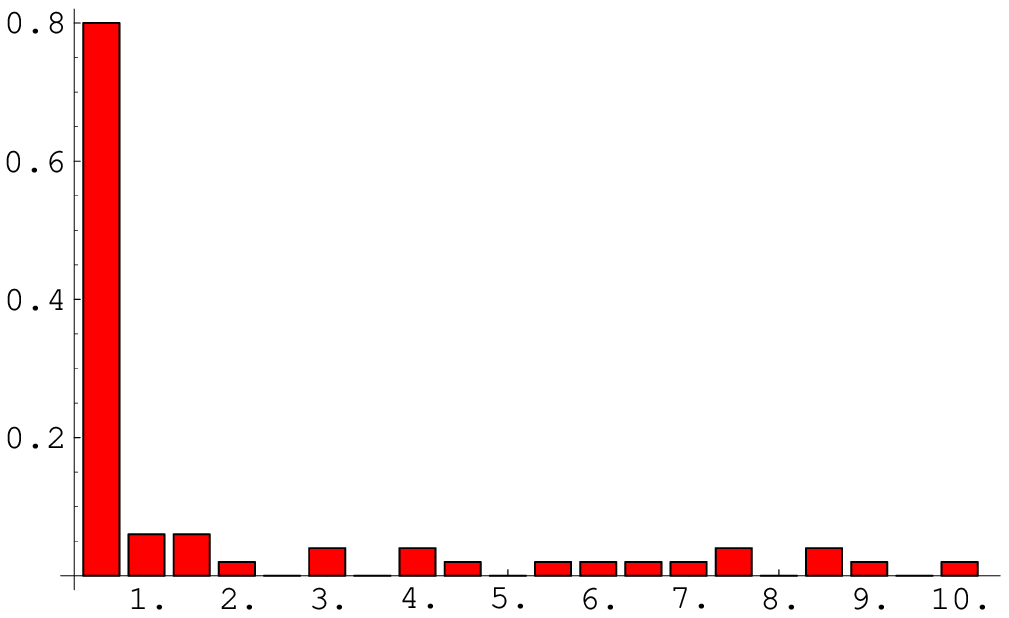}{.33} & \fig{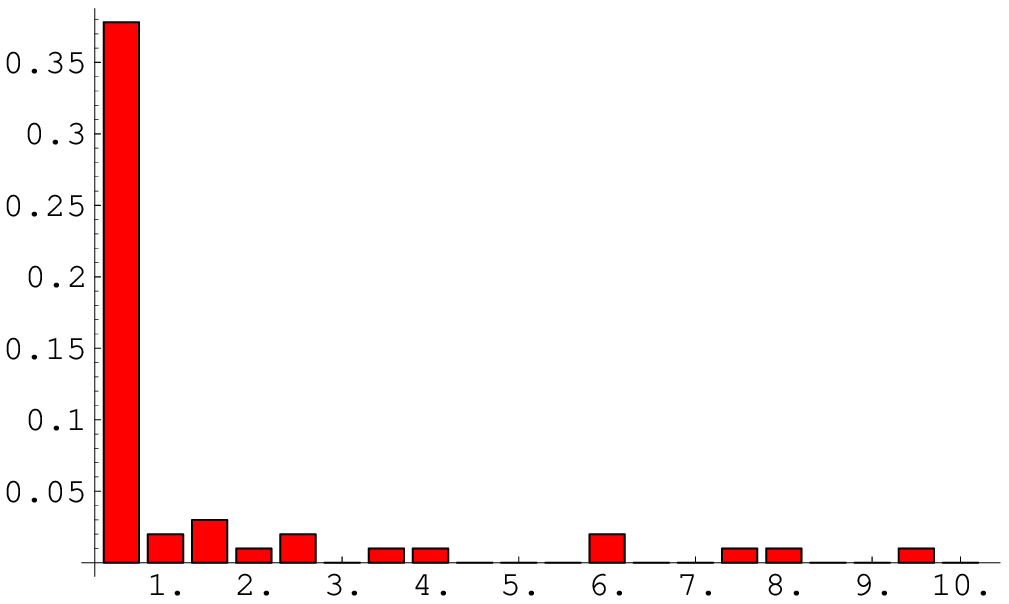}{.33} \\
$k=6.5$ and $p=1.5$ & \fig{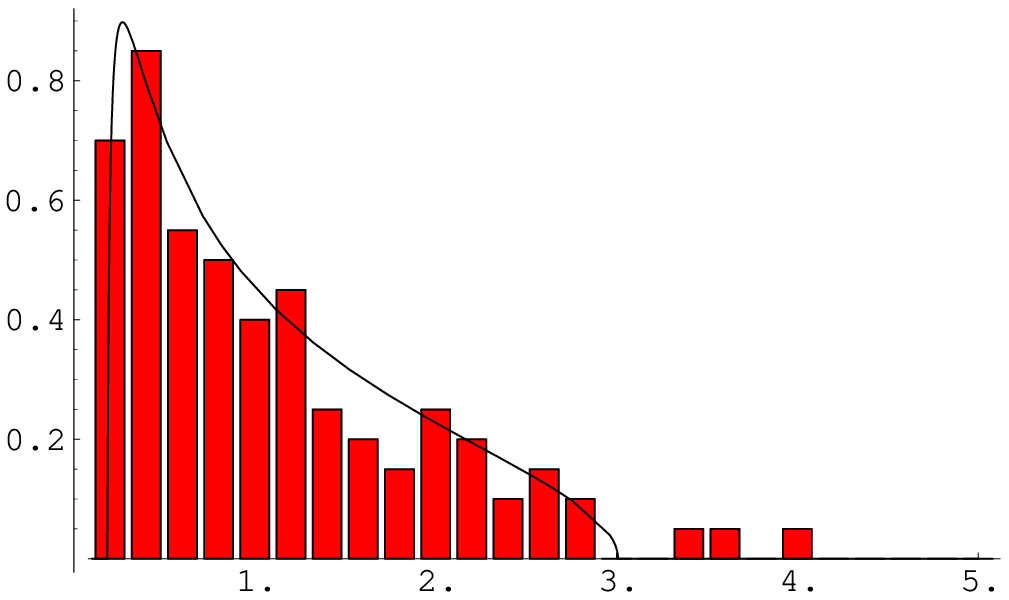}{.33} & \fig{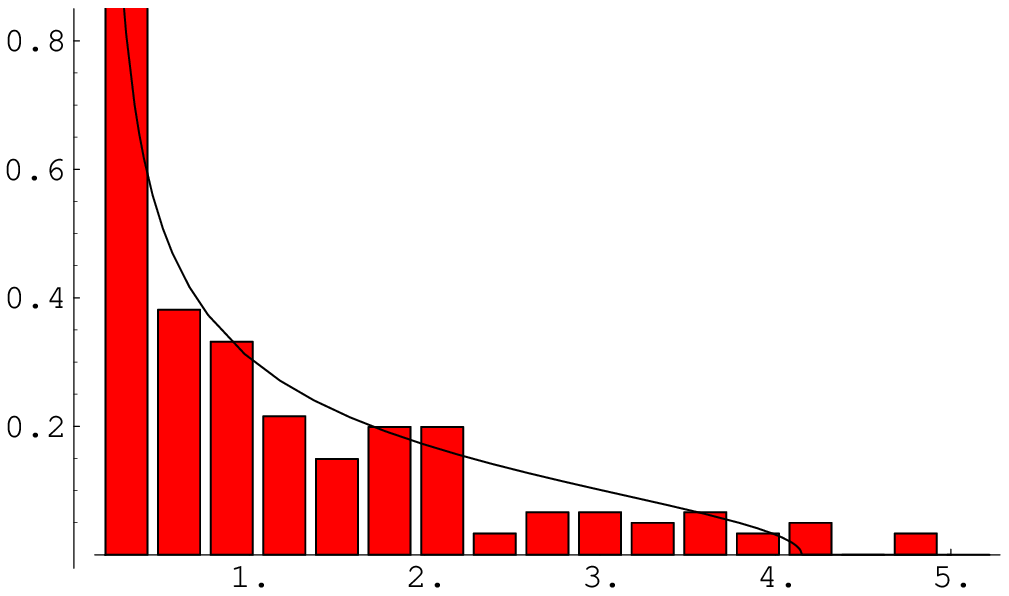}{.33} \\
\end{tabularx}
\caption{Eigenvalue density of Gram matrices for the quantum kicked top
for various values of $\tau,k$ en $p$ ($\tau$ is defined as the ratio
$K/N$ with $N=2j+1=201$)} 
\label{kickfig}
\end{figure*}

The same procedure was carried out for a second example: the quantum kicked
top.  This is a quantum spin-$j$ system with basic observables $J_x,J_y$ and
$J_z$ obeying the usual commutation relations $[J_x,J_y]=iJ_z\,  +\, cycl.\
perm.$.  The unitary evolution is given by
\begin{equation*}
 U = \mathrm{e}^{-\mathrm{i}kJ_z^2/2j} \mathrm{e}^{-\mathrm{i}pJ_y}.
\end{equation*}
The classical limit of this system is a dynamics on the two dimensional sphere
given by
\begin{align*}
 \hspace{-5mm} x' &= (x \cos p + z \sin p) \cos (k z \cos p - k x \sin p)\\
		  &\quad  + y \sin(k
 z \cos p - k x \sin p), \\
 \hspace{-5mm} y' &= (x \cos p + z \sin p) \sin (k z \cos p - k x \sin p)\\
		  &\quad - y \cos(k
 z \cos p - k x \sin p), \\
 \hspace{-5mm}z' &= - x \sin p + z \cos p,
\end{align*} 
with $(x,y,z)$ a unit vector in $\R^3$ which is the classical limit of
$(J_x,J_y,J_z)/j$. This map is a composition of a rotation around the
y-axis over an angle $p$ and a torsion around the $z$-axis over an
angle proportional to the $z$-component. Numerical studies suggest the following
behaviour. For small values of the parameters $k$ and $p$, this system displays
regular behaviour in large parts of the sphere, separated by chaotic regions.
The regular islands shrink as the parameters grow, leading eventually to
almost global chaos (See~\cite{HaKuSc}). This is an interesting feature of this
model: varying the control parameters $k$ and $p$ changes the behaviour of the
classical model considerably. It can be concluded from Figure~\ref{kickfig} that
this difference in behaviour is also present in the quantum kicked top. For
small values of the control parameters $k$ and $p$ there are a lot of
approximate returns during the first $K=\tau N $ time steps. This is clear from
the large number of eigenvalues close to zero.  We expect that in the limit
$K=\tau N\rightarrow \infty$ this distribution tends to a degenerate one in
zero as the system visits only a small fraction of the available states. Enlarging the control parameters places more weight on higher eigenvalues
and a totally different eigenvalue distribution appears. These distributions
appear to be very similar to the ones for the baker map for the same value of
$\tau$. 

One unexplained feature in the diagrams is the solid line. This is
actually the graph of an analytical function obtained in the following
way. Instead of trying to calculate the limiting empirical eigenvalue
distribution for a true quantum mechanical model, like the baker map or
the kicked top, one can try to simplify the calculations by considering
a random vector model. For an $N$-dimensional Hilbert space, a sequence
of $K$ normalised vectors is chosen randomly and independently
according to the uniform distribution on the unit sphere.  In this way,
one can try to mimic the evolution of a system which wanders wildly in
Hilbert space. The limiting eigenvalue distribution
($N,K\rightarrow\infty$ and $K/N\rightarrow\tau$) of the Gram matrix
associated with this sequence of random vectors can be calculated 
explicitly and is given by the Marchenko-Pastur
distribution~\cite{DeFaSp}
\begin{equation*}
 \hspace{-6mm}\rho_\tau(\r dt) = 
  \begin{cases}
   \displaystyle \frac{\sqrt{4\tau t - (t+\tau-1)^2}}{2\pi\tau t}\, \r dt
   & 0 < \tau \le 1 \\
   \displaystyle \frac{\tau-1}{\tau} \delta(t)\, \r dt \\
   \quad + \frac{\sqrt{4\tau t - (t+\tau-1)^2}}{2\pi\tau t}\,\r dt 
   & \qquad 1 < \tau. 
 \end{cases}
\end{equation*}
This density is the solid line depicted in Figures~\ref{bakerfig}
and~\ref{kickfig}. A possible explanation for the similarity between
the eigenvalue distributions generated by the baker  map and the kicked
top and the Marchenko-Pastur distribution is that, when the number of
time steps is linear in the dimension of the Hilbert space, subsequent
vectors in a chaotic quantum evolution are almost random and visit all
of the Hilbert space without preference for one direction or region.

\section{Conclusion}

The eigenvalue distribution of the Gram matrix appears to be a meaningful object
in the so-called deep quantum regime where the system is allowed to visit a
finite fraction of the total number of available states. Moreover, for systems
with a chaotic classical limit, numerical simulations indicate that the
limiting distribution is universal and coincides with that of random vectors,
even if the latter model is not a unitary dynamics. If the classical limit is
regular, we find a degenerate distribution. Linking the limiting spectrum of
the Gram matrix with the distribution of energy or quasi-energy level
separations~\cite{BeTa} definitely deserves further research. Another
interesting question, that is also relevant in quantum information theory, is
what happens when the dynamical system is no longer isolated but weakly coupled
to an environment. In such a situation, density matrices instead of vector
states must be used to describe the decoherence induced by the environment.

\end{document}